\def \yskip{\penalty-50\vskip3pt plus 3pt minus 2pt}
\def \reference{\par \yskip \noindent \hangindent .4in \hangafter 1}
\def \abc#1#2#3#4 {\reference#1, {\sl#2}, {\bf#3}, #4}
\def \blank {\lower 5pt\hbox to 0.75in{\hrulefill}}

\def \cm{~\rm{cm}}
\def \s{~\rm{s}}
\def \km{~\rm{km}}

\def \K{~\rm{K}}

\def \G{~\rm{G}}
\def \AU{~\rm{AU}}
\def \erg{~\rm{erg}}

\def \yr{~\rm{yr}}

\def \lesssim{\mathrel{<\kern-1.0em\lower0.9ex\hbox{$\sim$}}}
\def \gtrsim{\mathrel{>\kern-1.0em\lower0.9ex\hbox{$\sim$}}}

\documentstyle[11pt,aasms,tighten,flushrt]{article}
\begin{document}
%\normalsize
\small

\setcounter{page}{1}

%\shorttitle{CIRCUMSTELLAR MAGNETIC FIELDS}
%\shortauthors{SOKER}

\slugcomment{Draft version of \today}
%\slugcomment{Submitted to the Astrophysical Journal}

\title{LOCAL CIRCUMSTELLAR MAGNETIC FIELDS 
     AROUND EVOLVED STARS}

\author{Noam Soker}

\affil{Department of Physics, Oranim, Tivon 36006, Israel; 
soker@physics.technion.ac.il}
%Department of Astronomy, University of Virginia, P.O. Box 3818
%Charlottesville, VA 22903-0818;

%\altaffiltext{1}{On sabbatical from the University of Haifa at Oranim,
%Department of Physics, Oranim, Tivon 36006, Israel}

$$
$$

\centerline {\bf ABSTRACT}

I argue that the presence of magnetic fields around evolved stars,
e.g., asymptotic giant branch stars, and in PNe,
does not necessarily imply that the magnetic field plays a global
dynamical role in shaping the circumstellar envelope.
Instead, I favor magnetic fields with small coherence lengths,
which result from stellar magnetic spots or from jets blown by
an accreting companion.
Although the magnetic field does not play a global role
in shaping the circumstellar envelope, it may enhance local
motion (turbulent) via magnetic tension and reconnection.
The locally strong magnetic tension may enforce coherence flow
which may favor the masing process.

{\it Subject headings:} circumstellar matter
$-$ stars: magnetic fields
$-$ stars: AGB and post-AGB
$-$ masers
$-$ planetary nebulae: general

% ===================================================
\section{INTRODUCTION} 
% ===================================================

The axisymmetrical structure of most planetary
nebulae (PNe) motivated the development of models based on
magnetic fields.
The idea that the galactic magnetic field shapes PNe was raised
several times (e.g., Grinin \& Zvereva 1968), but was always
`killed', recently by Corradi, Aznar, \& Mampaso (1998).
More popular are models that attribute a dynamical role to the magnetic
field of the progenitor, during its asymptotic giant branch (AGB)
or post-AGB phases (e.g., Gurzadyan 1962; Woyk 1968;  Pascoli 1985, 1997;
Chevalier \& Luo 1994; Garc\'{\i}a-Segura 1997;
Garc\'{\i}a-Segura et al.\ 1999; Garc\'{\i}a-Segura, \& L\'opez 2000;
Garc\'{\i}a-Segura, L\'opez, \& Franco 2001; 
Matt et al.\ 2000; Blackman et al.\ 2001; Gardiner \& Frank 2001;
see review by Garc\'{\i}a-Segura 2002).
In several papers I expressed my objection to these models
(Soker \& Zoabi 2002 and references therein). 
There are now strong indications for the presence of magnetic
fields around AGB and other evolved stars and in PNe, mainly from
maser polarization; SiO maser close, typically less than a
few stellar radii, to the stellar surface (e.g. Kemball \& Diamond 1997),
and OH maser at $\sim 10^{15}-10^{16} \cm$ from the star (e.g., 
Zijlstra et al.\ 1989; Szymczak, Cohen, \& Richards 1999;
Miranda et al.\ 2001).
The polarization of SiO maser emission close to the stellar surface,
though, may result from anisotropy as a result of radiative pumping
and does not necessarily imply the presence of magnetic fields
(e.g., Bujarrabal, \& Nguyen-Q-Rieu 1981; Desmurs et al.\ 2000).
 
In a very interesting work, Miranda et al.\ (2001) find polarization
in the 1,665-MHz OH maser line, which indicates the presence of
$\sim 10^{-3} \G$ magnetic field at $\sim 10^{16} \cm$ from the
central star of the young PN K3-35.
They then argue, ``This result favours magnetic collimation
models of outflows in PNe.''
I find this statement unjustified.
To support my view, I discuss in the present paper other sources
of magnetic fields in PNe and AGB stars, which do not require the
magnetic field to play a globally important dynamical role.
In general, the other possible sources of the magnetic field are: 
\newline
(1) The magnetic field can be localized in specific regions.
This is predicted if there are stellar magnetic spots,
possibly with a higher mass loss rate from active regions, e.g.,
higher mass loss rate from cool magnetic spots,
which facilitate the formation of dust 
(e.g., Soker 1998; Soker \& Clayton 1999). 
In this process the magnetic field has a secondary role and
becomes dynamically important only in small local regions; 
the large scale magnetic field is too weak to play a dynamic role
and directly influence the wind from the AGB star.
This source of magnetic field close to the AGB star is discussed in $\S 2$.
\newline
(2) An accreting companion, whether a WD or a main sequence star, will
amplify the magnetic field. Hence the strong field may result
from an accreting companion.
The magnetic field will be carried out to the nebula by the jets
blown by the accreting companion.
These jets form the lobes in bipolar PNe (Morris 1987;
Soker \& Rappaport 2000).
This process may explain the presence of magnetic field in polar lobes
at large distances from the central star, 
e.g., as found by Miranda et al.\ (2001) in the PN K3-35.
This is discussed in $\S 3$.
\newline
(3) The magnetic field may result from a magnetically active main
sequence companion.
 Accreting main sequence companions will be spun-up and may become
magnetically active (Jeffries \& Stevens 1996; Soker \& Kastner 2002).
It is not clear how the magnetic field will be carried out
to the nebula in this case.
This source will not be discussed in the present paper.
        
% ===================================================
\section{THE MAGNETIC FIELD BEFORE DUST FORMATION} 
% ===================================================

\subsection {The Magnetic Field Intensity}

I consider a local magnetic field, e.g., such as above 
cool magnetic spots in AGB stars.
 Such a geometry requires much lower magnetic activity than
a strong magnetic field over the entire surface, as 
in the model of Hartquist \& Dayson (1997).
 In AGB stars the photosphere of cool magnetic spots is
above the rest of the photosphere (fig 1 of Soker \& Clayton 1999).
Just above the spot the magnetic field lines are radial, whereas to
the side of the spot the field lines are bent and become horizontal.
Let $r$ be the radial coordinate measured from the stellar center,
and let $y$ be the coordinate perpendicular to the radial direction
near the spot, i.e., parallel to the stellar surface, and measured from
the radial line through the center of the spot.
 The radial component of the magnetic field just above the spot center
and close to the surface is
\begin{equation}
B_{sr} = B_{s0} \left( \frac {r}{R_\ast} \right)^{-2}, \qquad {\rm for}
\qquad y \ll y_s,
\end{equation}
where $R_\ast$ is the stellar radius, $B_{s0}$ is the (radial)
magnetic field in the spot, and $y_s$ the radius of the spot on
the stellar surface.
For the horizontal (in the $y$ direction) magnetic field component
to the side and above the spot, I take a simple form
\begin{equation}
B_T = \eta B_{s0} \left( \frac {r}{R_\ast} \right)^{-\alpha}
\left( \frac {y}{y_s} \right)^{-2}.
\end{equation}
If the horizontal magnetic component results directly from the
spot at the same radial coordinate, then $\alpha=2$ as in equation (1),
whereas if the horizontal component results from wind expansion
from the stellar surface, then $\alpha =-1$, since the tangential
component of a magnetic field dragged with a radial wind decreases as
$r^{-1}$.
 The coefficient $\eta < 1$ incorporates the bending of the field
lines near the spot, from being radial to being horizontal (tangential).
Equations (1) and (2) are applicable to spots over the entire stellar
surface, whether the spots are near the poles or equator.

For comparison, the equatorial horizontal magnetic field 
that results from stellar rotation is given by (e.g.,
Chevalier \& Luo 1994)
\begin{equation}
B_T ({\rm equatorial}) =  B_{\rm sur} \frac {v_{\rm rot}}{v_w}
\left( \frac {r}{R_\ast} \right)^{-2}
\left( \frac {r}{R_\ast} -1 \right),
\end{equation}
 where $B_{\rm sur}$ is the (radial) magnetic field on the surface,
$v_{\rm rot}$ is the equatorial rotation velocity, and $v_w(r)$ the
wind velocity.
  Close to the stellar surface the wind material is still accelerated, and
the wind velocity can be quite low, and some material may even fall
back (e.g., Bobolotz, Diamond, \& Kamball 1997;
S\'anchez Contreras et al.\ 2002).
 I therefore scale the wind velocity in the SiO maser region with
$v_w = 1 \km \s^{-1}$.
  The equatorial magnetic field at $r=2 R_\ast$ due to rotation is
$B_T ({\rm equatorial}, r=2R_\ast) \simeq  0.25 B_{\rm sur}
(v_{\rm rot}/1 \km \s^{-1})$.
By comparing this with equation (2), I find that the amplification of
the tangential component due to rotation close to the star
becomes comparable to that near magnetic cool spots, for the same value of
the surface magnetic field, only for $v_{\rm rot} \gtrsim 1 \km \s^{-1}$.
 Such a rotation velocity implies that (i) the AGB star was spun up by
a stellar companion, via common envelope or tidal interaction
(Soker 2001), and (ii) the magnetic activity is most likely very strong
(Soker 2000) with the formation of many cool magnetic spots. 
 I therefore consider magnetic spots in this section.
 An interesting system in that respect is the symbiotic binary R Aquarii,
composed of a Mira variable and a hot companion, with an orbital period 
of $\sim 44 \yr$, and eccentricity $e \sim 0.8$, and in which the
maser shell seems to rotate around the Mira type star
(Hollis, et al.\ 2001).
The periastron separation is only $\sim 4 \AU$, and it is expected
that the companion had spun up the Mira type star.
 The maser shell is very clumpy and elongate (Hollis, et al.\ 2001).                    
 The clumpiness and non-spherical structure may hint at the
influence of magnetic spots in a fast rotating AGB star.

As noticed by Hartquist \& Dyson (1997), shocks formed by
stellar pulsation will amplify the tangential magnetic field.
In a strong adiabatic shock the density increases by a factor of 4,
and so does the transverse magnetic field; the magnetic pressure
increases by a factor of 16.
 The radiative cooling time is very short close to AGB stars.
 For a shock speed of $30 \km \s^{-1}$ the gas is shocked to a temperature
of $T > 10^4 \K$, from which the cooling time is
\begin{equation}
\tau_{\rm cool} \sim 500
\left( \frac {n}{3 \times 10^9 \cm^{-3}} \right)^{-1}
\left(\frac {T}{10^4 \K} \right)
\left(\frac {\Lambda}{10^{-23} \erg \cm^3 \s^{-1}} \right)^{-1} {\rm sec},
\end{equation}
where the cooling rate, energy per unit volume per unit time,
is given by $\Lambda n_e n_p$, and $n$, $n_e$, and $n_p$,
are the total, electron, and proton densities, respectively.
The cooling time is much shorter than the flow time over a typical
maser spot size $l_m$,
\begin{equation}
\tau_{\rm flow} \sim l_m/C_s = 10^6 
\left( \frac{l_m}{0.1 \AU} \right)
\left( \frac {c_s}{10 \km \s^{-1}} \right)^{-1} {\rm sec},
\end{equation}
where $c_s$ is the sound speed.
 The cooling of the post-shock gas results in further compression,
 with further amplification of the tangential component of the
 magnetic field.
  However, the magnetic field amplification is limited by the volume from
which magnetic field lines can be compressed together.
If the size in the radial direction of the region from which gas is
compressed is $\sim R_\ast$, and a typical maser spot size is
$l_m \sim 0.1 R_\ast$, then the maximum amplification of the tangential
magnetic field component is by a factor of $(R_\ast/l_m) \sim 10$. 
 Using equation (2) I find that the tangential component of
the magnetic field can be amplified to the original value of the
magnetic field in the cool magnetic spot in a region up to a tangential
distance from the cool spot of $\sim 2$ times the cool magnetic
spot radius.
 Closer to the spot the magnetic field will be stronger than
the original value.

 As discussed by Soker \& Clayton (1999; their $\S 4.1$), the
radial component of the magnetic field above the cool magnetic spot
is also amplified.
 The total pressure above the cool magnetic spot, which is the
 sum of the thermal and magnetic pressure, is equal to the
 pressure of its surroundings, which is mainly thermal pressure.
 After the passage of the shock front, the thermal pressure is
 increased by a large fraction, both above the cool spot and
 in its surroundings, while the radial magnetic field is not affected.
 As a result, the total pressure above the cool magnetic spot is lower
than that of the surrounding medium, hence the surrounding medium
compresses (in the tangential direction) the gas above the cool
spots and increases the intensity of the radial magnetic field.
This amplification is not as efficient as the amplification of the
tangential component, because the flow occurs after the shock passage,
and the amplified radial component is compressed to a small region
above the center of the spot. 
Despite that, the radial magneitc field component becomes the
dominate component above the center of a cool magnetic spot.
In the immediate surrounding of the magnetic spot on the stellar surface,
the tangential component of the magnetic field dominates in the
postshock region.
 The 90 degree sharp change in the direction of polarization at
several locations in TX cam (Kemball \& Diamond 1997),
may hint at the presence of radial magnetic field close to a
tangential field, hence to the presence of magnetic spots. 

To form a cool spot in AGB stars the photospheric magnetic pressure
should be of the order of the photospheric thermal pressure, which
gives $B_{s0} \simeq 10-100 \G$ on the photosphere
(Soker 1998).
This section shows that the magnetic field in the vicinity of cool
magnetic spots can reach a value close to that of $B_{s0}$;
tangential component can be amplified by a large factor, but the
coefficient $\eta$ in equation (1) is likely of order $\eta \sim 0.1$.
 I conclude that a magnetic field of $1-10 \G$ can be reached in
the vicinity of cool magnetic spots.
 Cool magnetic spots on the surface of AGB stars may have a radius of
$y_s \lesssim 0.1 R_\ast$ (Soker \& Clayton 1999).
Taking a typical radius of $y_s \simeq 0.05 R_\ast$ for the magnetic spot,
the size of maser spots with strong tangential
magnetic field can be $\sim 4 y_s \sim 0.2 R_\ast$,
increasing as $r$ away from the stellar surface. 
Even if the magnetic field is below, but not much below, its equipartition
value on the photosphere, i.e., $0.1 \G \lesssim B \lesssim 10 \G$,
it can still be amplified to become detectable.
Strong radial magnetic field regions will be above the center
of the cool spots and cover much smaller regions.
 This seems to be compatible with observations.

% ====================================================
\subsection {Dynamical Effects of the Magnetic Fields}
% ====================================================
 To shape the global flow by magnetic tension, the tangential component
of the magnetic field must circle the star (Chevalier \& Luo 1994;
Garc{\'{\i}}a-Segura 1997).
 This does not seem to be the case, at least in systems
where polarization changes direction at several locations,
e.g., in TX Cam.
 The magnetic tension, as expressed in units of force per unit volume,
is given by $f_{BT} = B^2/(4 \pi R_c)$, where $R_c$ is the curvature
radius of the magnetic field lines.
 In case of localized magnetic fields, e.g., magnetic field of
cool spots treated here, the radius of curvature is a few times
the size of the maser clumps.
 In TX Cam, for example, I find from Kemball \& Diamond (1997)
$R_c \sim 0.1-0.3 R_\ast$.
The acceleration of the wind extends over a distance of
$\gtrsim R_\ast$, which is the distance along which the relevant forces
change.
 Therefore, even if the the magnetic field is below the equipartition
(of pressure or energy) value, i.e., the magnetic pressure $P_B$ is
below the thermal pressure, $P_{\rm th}$, its {\it local} role can still
be significant if $P_B/R_c \gtrsim P_{\rm th}/R_\ast$.
 In particular, it may force a coherence flow of small parcels of gas,
in a way that favors the maser activity.
 If this is the case, maser clumps will tend to be located in regions
of strong magnetic fields.

 To conclude this subsection and the previous one, the presence of strong
magnetic fields in maser clumps close to AGB and red supergiant stars
does not imply at all that the magnetic field influence the flow on
a large scale, i.e., globally.
I argue that, more likely, these are local magnetic field structures,
formed above and near magnetic spots, which may
enhance the maser activity by enforcing coherent flow via magnetic
tension in these regions.
 Magnetic tension by magnetic field lines which are anchord to the
stellar surface, may even cause some material to fall back,
similar to such a flow which was suggested by Wang \& Sheeley (2002)
to occur above the solar photosphere.
  
% ===================================================
\section{AMPLIFICATION AT LARGE DISTANCES} 
% ===================================================
At large distances from the star in radially expanding winds,
tangential component of the magnetic field dominates,
and the magnetic pressure and energy
density decrease as $r^{-2}$, like the wind's kinetic
energy density (e.g., Chevalier \& Luo 1994).
Hence, the ratio of the magnetic energy density to kinetic energy
density, $\sigma$, stays constant for a freely expanding wind.
 This ratio, or the ratio of magnetic to thermal energy, increases
as the wind slows down via a shock wave or in the postshock region.
 The compression of the wind in the radial direction that increases
the density by a ratio $\Gamma$, increases the tangential
magnetic field component by the same factor, hence the magnetic
energy density by a factor $\Gamma^2$. 
Hence, in an isothermal flow the ratio $\sigma$ increases by
a factor of $\sim \Gamma$.
In a case of radiative cooling, the ratio of the the magnetic
energy density to that of thermal energy density can increase
by a much larger factor.
 
Radiative cooling is important at high densities, which
for expanding winds occur close to the central star.
Interaction of two winds close to the central star takes place
when the two winds are blown simultaneously, meaning one
wind, the slow one, is blown by the AGB (or post-AGB) progenitor,
and a fast wind, most likely collimated fast wind (CFW), is blown
by an accreting companion.
Such a situation occurs in symbiotic stars.
Indeed, Crocker et al.\ (2001) propose, based on their study of the
non-thermal emission of the symbiotic star CH Cygni, that
the interaction of jets with the circumstellar wind leads to local
enhancement in the magnetic field.
For typical parameters in binary progenitors of bipolar PNe,
radiative cooling may be significant at distances $\lesssim 10^{16} \cm$
from the central binary system (Soker 2002).
Fast wind from the central star in the post-AGB phase, which
blows after the slow wind ceases, can
have the same effect in principle.
The shocked material, being that of the slow wind or the CFW,
is compressed to a density where its thermal pressure,
$\rho_c c_c^2$, equals the
ram pressure of the CFW that forms the bipolar lobes,
$\rho_j v_j^2$.
Here $\rho_c$ is the density in the compressed shell,
$c_c$ the isothermal sound speed in this shell, and
$\rho_j$ and $v_j$ the density and expansion velocity of the jets
(or  CFW).
If each of the two jets expand into a solid angle $\Omega = 4 \pi \beta$,
and the mass loss rate (defined positively) into each jet is $\dot M_j$,
then $\rho_j = \dot M_j / (4 \pi \beta v_j r^2)$.
 The density of the freely expanding spherically symmetric slow wind
is $\rho_s = \dot M_s / (4 \pi v_s r^2)$,
where $\dot M_s$ and $v_s$ are the mass loss rate and velocity of
the slow wind.
 Equating the pressures then, gives the compression ratio of
the slow wind
\begin{equation}
\frac {\rho_c}{\rho_s} =
\left( \frac{v_s^2}{c_c^2} \right)
\left( \frac{\dot M_j v_j}{\dot M_s v_s} \right)
\frac{1}{\beta}.
\end{equation}
For typical parameters in binary systems forming lobes, the second
parenthesis is $\sim 0.1$, $\beta \sim 0.1-0.001$ (Soker 2002),
$v_s \simeq 10 \km \s^{-1}$, and before ionization starts in the
PN phase, $c_c \sim 1 \km \s^{-1}$.
 I find, therefore, that compression by a factor of $\sim 10^2$
by jets (or CFW) blown by a companion is plausible during
the AGB and proto-PN phases.

Therefore, I propose that the magnetic field found by Miranda
et al.\ (2001) along the polar directions of K3-35, was locally
amplified via the compression of material by propagating jets
blown by an accreting companion.
The bipolar point-symmetric structure of K3-35 suggests that it
was shaped by such jets.
The compressed material and magnetic field can originate
in the slow wind and/or in the jets.
The magnetic field does not have global dynamical effects,
but if it results from the jets,
it may have a coherence length similar in size to that of the cross
section of the lobes.

There is another effect of the CFW model.
 A CFW (or jets, one at each side of the equatorial plane)
will stretch magnetic field lines in the radial direction.
A bubble that is formed from the shocked slow wind
material and/or the CFW material will form a dense shell
on the boundary of the bubble, where the magnetic field lines
are compressed and amplified.
 Hence, in the dense shell on the sides of a lobe
strong radial magnetic field can be formed.
This is an alternative explanation, as compared with
a dipole field (Szymczak, Cohen, \& Richards 2001b), to the presence of
strong radial magnetic field lines at $\sim 2 \times 10^{16} \cm$
from the supergiant star VX Sgr (Szymczak et al.\ 2001b).
At a distance of 300 times the stellar radius the radial
component of the magnetic field is $\sim 10^{-5}$ that at
the surface of the star.
 If the magnetic field is compressed by a factor of $\sim 100$,
the new radial magnetic component is $\sim 10^{-3}$ its
surface value.
For a stellar surface value of $\sim 1 \G$, a radial magnetic field
of $\sim 10^{-3} \G$ is obtained.
Even higher values can be obtained if the tangential component,
which decreases with density only as $r^{-1}$, is stretched to
form the radial magnetic field before compression to the sides of
the lobe.

I argued above  that the magnetic field has no global dynamical effects.
However, as is the case close to the stellar surface studied in the
previous section, it does have local dynamical effects.
In particular, magnetic field reconnection can take place, leading
to fast flow, of the order of the Alfven speed, from the
reconnection site.
Bode \& Kahn (1985) propose that magnetic field reconnection
occurs in the wind of the recurrent nova RS Oph, a flow which resemble
in some sense the situation studied here.
The flow near the reconnection sites can lead to fast variations
in location and velocity of maser spots.
Palen \& Fix (2000) propose local effects of magnetic field
to account for the fast velocity variations in the OH
maser components in U Herculis.
They propose a different mechanism than reconnection, involving
turbulence operating with magnetic fields.
In addition to reconnection, the small coherence length of the
magnetic field proposed here implies that as the magnetic field
is amplified, magnetic tension can become strong enough to
accelerate material locally and enhance turbulent motion.

% ===================================================
\section{SUMMARY} 
% ===================================================

The single goal of this paper is to argue that the presence of
magnetic fields around evolved stars, e.g., AGB stars, and in PNe,
does not necessarily imply that the magnetic field plays a global
dynamical role in the shaping of the circumstellar matter.
This is in dispute with claims made in several papers in recent years,
e.g., Miranda et al.\ (2001).
I argued above that in most cases the magnetic field observed around
evolved stars and in PNe, are likely to originate from magnetic
spots on the stellar progenitor surface, or from jets blown by
an accreting companion.
In the later case strong magnetic fields will be formed on the
boundary of the lobes formed by the two jets.
I favor this explanation to the magnetic fields found by
Miranda et al.\ (2001) in K3-35. 

The following comments are in place here.
\newline
(1) In both cases for the origin of the magnetic field, the magnetic
field has a short coherence length, such that magnetic
tension may become locally important.
Hence, although the magnetic field does not play a global role
in shaping the circumstellar matter, it may enhance turbulent
motion via magnetic tension and reconnection.
\newline
(2)  In cases when the evolved star rotates fast enough,
the dynamo activity may become very efficient, such that the
magnetic field does have a global dynamical role.
This may occur for massive stars before they lose most of their envelope
(but not stars on the upper AGB that have lost most of their envelope),
and stars that have been spun up by a stellar companion
(Soker 2001).
\newline
(3) Magnetic spots, as in the sun, are likely to possess a global
polarity. Hence, a general global polarization sense may be detected
in masers, even if there are fluctuations on much smaller scale
and the magnetic field has no global role.
Such a structure is inferred in some objects 
(e.g., Szymczak et al.\ 1999, 2001a).
\newline
(4) The locally strong magnetic tension may enforce coherence flow
that may favor the masing process.
Local magnetic tension above the stellar surface may pull material back
to the surface, similar to such a flow above the solar surface
(Wang \& Sheeley 2002).

\acknowledgements
Part of this work was done at the University of Virginia, where
I was supported by a Celerity Foundation Distinguished Visiting 
Scholar grant.
 This research was supported by the US-Israel Binational Science Foundation.

\end{document}